\documentclass[aps,preprint,showpacs,amsmath,amssymb]{revtex4}
\usepackage{epsfig}
\usepackage{wasysym}
\usepackage{amssymb}

\begin{document}

%\tighten
\newcommand{\bvec}[1]{\mbox{\boldmath ${#1}$}}
%\newcommand{\bm}{\boldmath}
%\preprint{\vbox{ \hfill FIS-UI-TH-XXXX}}
\title{Constraining the mass and width of the $N^*(1685)$ resonance}
\author{T. Mart}
\affiliation{Departemen Fisika, FMIPA, Universitas Indonesia, Depok 16424, 
  Indonesia}
\date{\today}
\begin{abstract}
 We have examined the existence of the $N^*(1685)$ resonance, which is
 recently listed by the Particle Data Group as a one-star nucleon
 resonance, by using a covariant isobar model for kaon photoproduction 
 and assuming that the resonance
 has $J^p=1/2^+$, in accordance with our previous finding. After the 
 inclusion of this resonance the changes in the total $\chi^2$ show two
 clear minima at $M_{N^*}=1650$ and 1696 MeV, which correspond to
 two different resonance states. The former corresponds
 to the narrow nucleon resonance found in our previous investigation,
 whereas the latter corresponds to a new resonance found as
 we increase the resonance width.  From the latter we
 derive the mass and width relation of the $N^*(1685)$ resonance. 
 We observe that the properties of both 
 the $N^*(1685)$ and $N^*(1710)P_{11}$ resonances are strongly correlated.
 Although the best fit of the present work
 yields $M_{N^*}=1696$ MeV and
 $\Gamma_{N^*}=76$ MeV, the apparently 
 small $N^*(1710)P_{11}$ coupling constant to the $K^+\Lambda$ channel 
 found in previous investigations suggests that $\Gamma_{N^*}\lesssim 35$ MeV, which,
 according to the mass and width relation, corresponds to 
 $M_{N^*}\lesssim 1680$ MeV. 
\end{abstract}

\maketitle
%\newpage

In a previous work \cite{Mart:2011ey} 
we have investigated the possibility
of observing the $J^p=1/2^+$ narrow resonance predicted by the 
chiral soliton model as a member of baryon anti-decuplet 
\cite{diakonov} by using the 
kaon photoproduction process $\gamma +p\to K^++\Lambda$.
For this purpose an isobar model, for which the background
part is constructed from a covariant diagrammatic 
technique and the resonance part is written in terms of
the electric and magnetic multipoles, was fitted to the
low energy (near threshold) photoproduction data. 
A new narrow resonance was added to the model and the 
effect of the resonance on reducing the $\chi^2$ was
investigated by scanning its mass from 1620 to 1730 MeV.
By varying the total width from 0.1 to 10 MeV it was 
found that the most promising mass of the resonance is
1650 MeV, whereas the corresponding width is 5 MeV.
The possibility that the resonance has different quantum
numbers has been also investigated. Nevertheless, 
at present, experimental 
data indicate that the  $J^p=1/2^+$
is the most suitable quantum number for this 
state \cite{Mart:2011ey}.

Recently, the $N^*(1685)$ resonance has been listed
by the Particle Data Group (PDG) as a new state with one-star
status in the 2012 Review of Particle Physics 
\cite{Beringer:1900zz}. Its spin and parity are still
undetermined. The quoted literatures originate
from the recent experimental measurements of the 
$\eta$ photoproduction on a neutron by the  
CBELSA/TAPS Collaborations \cite{Jaegle:2008ux,Jaegle:2011sw} 
and the quasi-free Compton scattering on the neutron
by the GRAAL Collaboration \cite{Kuznetsov:2010as}.
The resonance mass and width extracted from the 
CBELSA/TAPS experiment  are  1670 and 25 MeV,
respectively, whereas those from GRAAL data  are 
1685 and $\le$30 MeV, respectively.

Interestingly, however, in the 2012 PDG list of 
the $N^*(1685)$ it is noted that this state does 
not gain status by being a sought-after member 
of a baryon anti-decuplet \cite{Beringer:1900zz}.
Therefore, the $N^*(1685)$ state should not be
considered merely as a member of an anti-decuplet
narrow resonance, it could belong to the family of  
usual nucleon resonances in the PDG listings. 

In view of this it is obviously important to
relax the upper limit of the resonance width in
our previous investigation  \cite{Mart:2011ey}. 
Furthermore, our previous finding reveals not only
one possible resonance mass at 1650 MeV, but also
three other minima at 1680, 1700, and 1720 MeV, albeit 
with weaker signals compared to that at 1650 MeV.

In the present study we use our latest isobar model 
constructed from appropriate Feynman diagrams  for
both background and resonance parts \cite{mart2012}. 
The model fits all latest available $K^+\Lambda$ photoproduction data
up to $W=2200$ MeV, 
consisting of differential cross section, recoil polarization,
beam-recoil double polarization, as well as photon $\Sigma$ 
and target $T$ asymmetries data. In total, the 
fitting database consists of more than 3500 data points. 
A more detailed explanation of the model as well as 
the experimental data used to fit the model 
can be found in Ref.~\cite{mart2012}.
Therefore, the presently used isobar model is a fully covariant
model and fits not only the experimental data near threshold.
The use of this model has certain advantages 
compared to the previous one, e.g., it is clearly safe 
to extend the investigation beyond the threshold energy,  
and it also provides a good tool for cross checking  
our previous result \cite{Mart:2011ey}.

\begin{figure}[t]
  \begin{center}
    \leavevmode
    \epsfig{figure=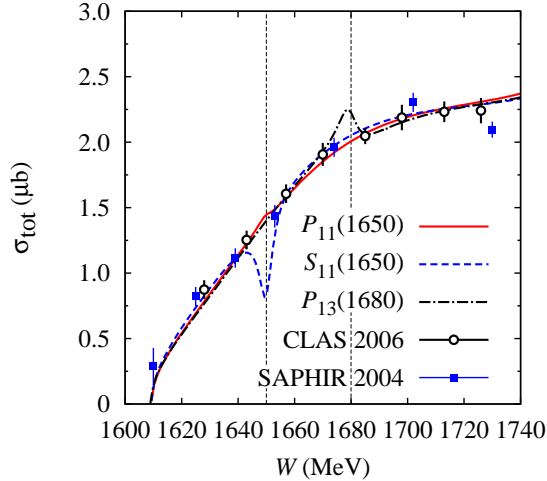,width=75mm}
    \caption{(Color online) Effects of the inclusion of 
	$P_{11}$, $S_{11}$, and $P_{13}$  resonances on the total cross 
	section of the $\gamma+p\to{K^+}+\Lambda$ process in our
        previous calculation \cite{Mart:2011ey}. Note that the 
        $P_{13}(1680)$ effect was not reported in Ref.~\cite{Mart:2011ey}.
        Experimental data are from the SAPHIR \cite{Glander:2003jw}
        and  CLAS \cite{Bradford:2005pt} collaborations.
      }
   \label{fig:kltot_hnp} 
  \end{center}
\end{figure}

We assume that the state has quantum numbers $J^p=1/2^+$,
i.e., the $P_{11}$,  since our previous study found 
that these quantum numbers are the most suitable one
for investigation using kaon photoproduction. 
The state $J^p=1/2^-$, for instance, can be 
ruled out by the present experimental data because 
it existence generates a clear dip at
$W=1650$ MeV in the total and differential cross sections,
which are experimentally not observed (see Fig.~\ref{fig:kltot_hnp}
for the total cross section case as well as Figs. 14 and 15 of 
Ref.~\cite{Mart:2011ey} for a more detailed result).
The use of total cross section is necessary at this stage 
because the effect is small in the differential cross
section, whereas in the total cross section  the cumulative effect
could be larger due to a constructive interference in the
$S_{11}$ case or even much smaller due to a destructive one
in the $P_{11}$ case (see Fig.~\ref{fig:kltot_hnp}).

Investigation of the higher spin resonance is somewhat problematic in
the current method. As discussed in Ref.~\cite{Mart:2011ey}, for instance,
the use of $P_{13}$ state results in weaker and
more complicated changes in 
the total $\chi^2$. In fact, as shown in Fig. 13 of Ref.~\cite{Mart:2011ey},
it replaces the strong signal at 1650 MeV with a
rather weaker one at 1680 MeV. 
It should be emphasized here that the signal of
the narrow resonance at  1650 MeV was obtained by using
the $S_{11}$ or $P_{11}$ state, independent of the 
$K\Lambda$ branching ratio and the total decay width used. 
There are two
possible explanations of this phenomenon. First, it shows the
limitation of the present approach  for investigating
the higher spin states due to their complicated structures.
Second, the presently available experimental data are 
not sufficiently accurate for this purpose. For the latter,
future kaon photoproduction experiments at MAMI, Mainz, could
be expected to overcome this problem. Furthermore, we also observe
that the use of the  $P_{13}(1680)$ state leads to a clear bump at 
$W=1680$ MeV in the total cross section, as shown in 
Fig.~\ref{fig:kltot_hnp}, that is also not observed
by the presently available experimental data. 

\begin{figure}[t]
  \begin{center}
    \leavevmode
    \epsfig{figure=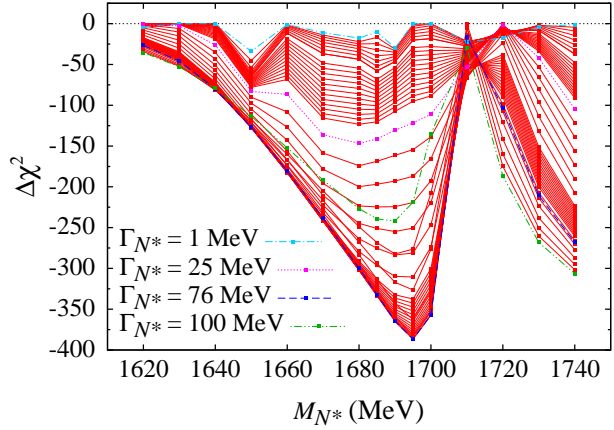,width=85mm}
    \caption{(Color online) Changes of the $\chi^2$ in the
	fit of the kaon photoproduction data
        due to the inclusion of the $N^*(1685)P_{11}$ 
	resonance with the corresponding mass 
        and width scanned from 1620 
	to 1730 MeV and  1 to 100 MeV, respectively.
      }
   \label{fig:scan-mass} 
  \end{center}
\end{figure}

As in the previous study we perform fits by adjusting the whole
unknown coupling constants in the model and scan the resonance
mass from 1620 to 1740 MeV, with 10 MeV step, and calculate
the changes in the total $\chi^2$ after including the $P_{11}$ 
resonance. Near the global minimum
we decrease the step to 5 MeV in order to increase the accuracy.
However, different from the previous study, here we vary the
total width from 1 to 100 MeV. The result is displayed in
Fig.~\ref{fig:scan-mass}. In total, we have performed 780
fits to produce this figure. Figure~\ref{fig:scan-mass} 
clearly exhibits that we recover the finding in
our previous study \cite{Mart:2011ey}, 
i.e.,  the 1650 MeV narrow resonance, which 
is indicated by the first minimum at $M_{N^*}=1650$ MeV.
This provides a proof that our present 
covariant isobar model is consistent with
the model used in the previous study. 
Surprisingly, however, this minimum only exists for the total width
$\Gamma_{N^*} \le 25$ MeV. As the total width increases beyond
this value, the minimum at 1650 MeV gradually vanishes.
This fact was not observed in our previous study because
the width was limited only up to 10 MeV.
Thus, we believe that our present finding still supports
the existence of a narrow resonance at 1650 MeV. Although
the width could be larger, the upper limit of 25 MeV obviously
exhibits that the  corresponding resonance remains narrow.
The second (global) minimum is our best fit and 
found at $M_{N^*}=1696$ MeV 
with the corresponding
width $\Gamma_{N^*}=76$ MeV (see Fig.~\ref{fig:scan-mass}). 
There is a sign of another minimum for
$M_{N^*}> 1740$ MeV, which is, however, beyond our present interest.

\begin{figure}[t]
  \begin{center}
    \leavevmode
    \epsfig{figure=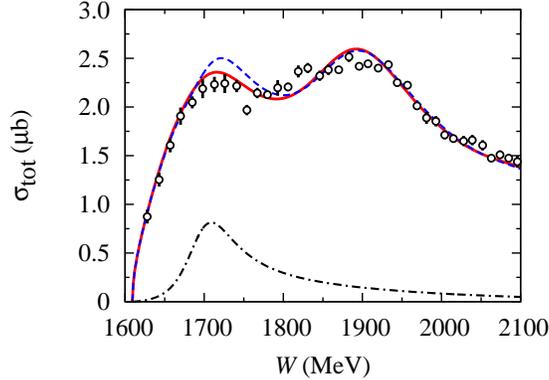,width=75mm}
    \caption{(Color online) Total cross sections obtained by the model calculations 
      with (solid red line) and without (dashed blue line) the
      $N^*(1685)P_{11}$ resonance compared with experimental
      data \cite{Bradford:2005pt}. In the former, the total
      cross section is calculated by using the best fit result,
      i.e., with $M_{N^*}=1696$ MeV and $\Gamma_{N^*}=76$ MeV.
      The dash-dotted black line exhibits 
      the $N^*(1685)P_{11}$ resonance contribution. 
      }
   \label{fig:kltot} 
  \end{center}
\end{figure}

Figure \ref{fig:kltot} graphically displays the effect of  
inclusion of the $N^*(1685)P_{11}$ resonance on the improvement of 
the model. Note that the choice of the total cross section here is 
trivial and only
for the sake of convenience, because improvements after including
this resonance are not only found in cross sections,
but also in polarization observables. Moreover, the total cross
section data shown in Fig.~\ref{fig:kltot}
were not included during the fitting process. 

Obviously, the $N^*(1685)P_{11}$ resonance is important 
in improving the agreement between
experimental data and model calculation at $W\approx 1700$ MeV.
However, there might be a question regarding the large contribution
of the  $N^*(1685)P_{11}$ resonance as shown by the dash-dotted
line in Fig.~\ref{fig:kltot}, whereas on the other hand the effect
in decreasing the cross section at $W\approx 1700$ MeV seems to be
relatively small. This originates from the fact that in obtaining the
solid line [model with the  $N^*(1685)P_{11}$ resonance]
all background and resonance coupling constants in the model
are refitted. In fact, only by a destructive interference with the
contribution of the $N^*(1710)P_{11}$ resonance
the large $N^*(1685)P_{11}$ contribution can be 
compensated. We will discuss this problem later, when
we explain the relation between the mass and the total width
of the resonance.

\begin{figure}[t]
  \begin{center}
    \leavevmode
    \epsfig{figure=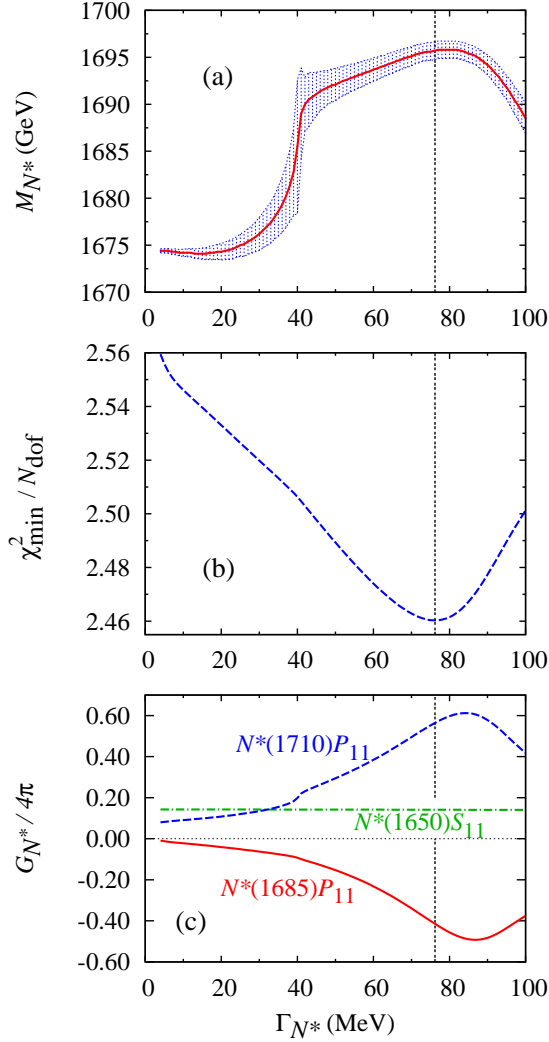,width=75mm}
    \caption{(Color online) (a) Mass and width relation of the $N^*(1685)P_{11}$ 
      resonance extracted from
      Fig.~\ref{fig:scan-mass} (solid red line). The uncertainty of this
      relation is indicated by the dotted (blue) area.
      (b) $\chi^2_{\rm min}$ per number of 
      degrees of freedom ($N_{\rm dof}$) 
      obtained for each of the $\Gamma_{N^*}$ values. The vertical
      dotted line locates the position of the 
      minimum value of the $\chi^2_{\rm min}/N_{\rm dof}$.
      (c) Coupling constants of the dominant resonances that
      contribute to the first peak of the $\gamma p\to K^+\Lambda$ 
      total cross section (see Fig.~\ref{fig:kltot}). Notation
      of the coupling constants can be found in Ref.~\cite{mart2012}.
      }
   \label{fig:scan-width} 
  \end{center}
\end{figure}

There is another interesting phenomenon revealed by 
Fig.~\ref{fig:scan-mass}, i.e., the position of the
second minimum varies as functions of the mass and
width of the resonance. The four different lines
for $\Gamma_{N^*}=1,25,76$ and 100 MeV in this
figure clearly suggests that there is a unique
relation between the mass and the width of the
resonance which can be obtained by fixing the
value of the total width $\Gamma_{N^*}$ during the fitting process
and finding the minimum value of $\chi^2$ per number of 
degrees of freedom, which is denoted from now on 
by $\chi^2_{\rm min}/N_{\rm dof}$.
By repeating this procedure for $\Gamma_{N^*}=1,\dots, 100$ MeV,
within the range of $M_{N^*}=1660-1710$ MeV in order
to exclude other minima, 
we obtain this relation which is plotted by using solid (red) line
in the panel (a) of 
Fig.~\ref{fig:scan-width}. Note that the uncertainty of
this relation, displayed by the shaded (blue) area, is obtained
from the fitted mass uncertainty given by MINUIT. 

In Fig.~\ref{fig:scan-width} all curves are
started from 4 MeV, since below this point the minimum mass is
immediately shifted to 1690 MeV, causing a discontinuity
in the plot. We may consider this minimum
as another case of narrow resonance, the same as the one
we found at 1650 MeV. As 
a consequence, it does not belong to the same second minimum.
In our previous study \cite{Mart:2011ey} we carefully stated
that around $W\approx 1690$ MeV there are threshold energies
of the $K^+\Sigma^0$, $K^+\Sigma^-$, $K^0\Sigma^+$ and $K^0\Sigma^0$
channels. This could provide an alternative explanation of this 
minimum, although experimental cross section data do not
show any discontinuity at this energy point.

The $\chi^2_{\rm min}/N_{\rm dof}$ plotted in the panel (b) 
of  Fig.~\ref{fig:scan-width} 
obviously advocates that
$M_{N^*}=1696$ MeV and $\Gamma_{N^*}=76$ MeV would be the
best result of the present work, 
consistent with that obtained from
Fig.~\ref{fig:scan-mass}. They are, however, substantially larger than 
those found in the latest $\eta$ photoproduction experiment 
($M_{N^*}=1670$ MeV and $\Gamma_{N^*}=25$ MeV) 
\cite{Jaegle:2008ux,Jaegle:2011sw}, 
as well as those extracted from 
the quasi-free Compton scattering on the neutron 
($M_{N^*}=1685$ MeV and $\Gamma_{N^*}\le 30$ MeV) \cite{Kuznetsov:2010as}.
These discrepancies may raise a serious problem.
However, the minimum value of $\chi^2_{\rm min}/N_{\rm dof}$ 
is not by any means the only criterion for a best result. 

We note that if we could fix the width of the resonance
to 25 MeV, the mass and width relation would yield $M_{N^*}\approx 1675$ MeV,
very close to that obtained from the recent $\eta$ photoproduction
experiment \cite{Jaegle:2008ux,Jaegle:2011sw}.
On the other hand, we also observe that our best fit 
($M_{N^*}=1696$ MeV and $\Gamma_{N^*}=76$ MeV) 
is obtained by increasing the contribution of the
$N^*(1685)P_{11}$ resonance. This is shown in the panel (c)
of Fig.~\ref{fig:scan-width}, where we can see that the absolute
value of the coupling constant increases as we increase $\Gamma_{N^*}$
up to $\sim 85$ MeV, whereas the $N^*(1650)S_{11}$ coupling 
is almost unaffected. To reproduce the experimental data
this process must be followed by increasing the
$N^*(1710)P_{11}$ coupling constant, where the coupling sign
should be different in order to produce a destructive interference
between the $N^*(1685)P_{11}$ and $N^*(1710)P_{11}$ amplitudes.

Previous studies \cite{igor,Mart:2006dk,mart2012,maxwell} 
have indicated that the $N^*(1710)P_{11}$ contribution to 
kaon photoproduction tends to be small. For instance, our previous
study \cite{mart2012} yields $G_{N^*(1710)}/4\pi\approx 0.08$, 
smaller than that of the $N^*(1650)S_{11}$,
i.e., $G_{N^*(1650)}/4\pi\approx 0.14$ . The small $N^*(1710)P_{11}$ 
contribution to the kaon photo- and electroproduction has been also pointed
out in a recent investigation by Maxwell \cite{maxwell}.
Although Maxwell found that both $N^*(1650)S_{11}$ and $N^*(1710)P_{11}$
resonances have comparable contributions, they are relatively smaller than
those of other nucleon resonances. A recent Bayesian analysis of the
$K^+\Lambda$ photoproduction has also 
excluded the $N^*(1710)P_{11}$ from
the list of resonances that have the highest probability of 
contributing to the reaction \cite{DeCruz:2011xi}.
Finally, the latest
GWU analysis has found no evidence for the $N^*(1710)P_{11}$ state,
in contrast to the $N^*(1650)S_{11}$ and $N^*(1720)P_{13}$
resonances \cite{arndt2006}, although the Bonn-Gatchina
group \cite{anisovich-EPJA} could draw a different conclusion.
 
Therefore, by referring to 
the  panel (c) of Fig.~\ref{fig:scan-width} and 
considering the fact that the $N^*(1710)P_{11}$  contribution
should be smaller than, or at least comparable to, 
the $N^*(1650)S_{11}$ contribution, 
we may estimate that the largest,
but still reasonable, 
total width of the $N^*(1710)P_{11}$ resonance is 
$\Gamma_{N^*}\approx 35$ MeV, for which the $N^*(1710)P_{11}$ and
$N^*(1650)S_{11}$ contributions are almost equal. 
According to the mass and width relation given in the panel (a) 
of Fig.~\ref{fig:scan-width} this result corresponds to 
$M_{N^*}\approx 1680$ MeV, which is certainly consistent with the recent
PDG estimate \cite{Beringer:1900zz}.

In conclusion, we have investigated the existence of the $N^*(1685)$ resonance
by using a covariant isobar model and calculating the changes in the total 
$\chi^2$ after including this  resonance. We assume that the resonance has
$J^p=1/2^+$, in accordance with our previous finding of the narrow resonance
in kaon photoproduction. Two clear signals are found at
$M_{N^*}=1650$ and 1696 MeV, which correspond to two different
nucleon resonances.
The former is the same narrow resonance found in our
previous study and only exists for $\Gamma_{N^*}\le 25$ MeV, whereas 
the latter is related to the best $\chi^2_{\rm min}$ found 
in the present calculation and is obtained with $\Gamma_{N^*}= 76$ MeV. 
From the changes in the total $\chi^2$ 
we have derived the relation between the mass and width of the resonance.
Our finding indicates that the properties of the 
$N^*(1685)$ and $N^*(1710)$ resonances are strongly correlated, since
the best $\chi^2_{\rm min}$ would be obtained
by increasing the contributions of both $N^*(1685)$ and $N^*(1710)$ resonances,
albeit with different coupling signs, in order to produce the
required destructive
interference. By considering the $N^*(1710)$ coupling found in previous
investigations we estimate the largest resonance width to be 
35 MeV, with the corresponding mass 1680 MeV. However, if we used
the width obtained from the latest $\eta$ photoproduction experiment,
i.e., 25 MeV, the resonance mass would be 1675 MeV,
consistent with the recent PDG estimate.

The author acknowledges support from the University of Indonesia.

\end{document}